\documentclass[aps,onecolumn]{revtex4}
\usepackage{epsfig,epsf}
\usepackage{graphicx}
\newcommand{\be}{\begin{equation}}
\newcommand{\ee}{\end{equation}}
\newcommand\beq{\begin{eqnarray}}
\newcommand\eeq{\end{eqnarray}}
\begin{document}

\title{Non-uniform chiral and 2SC color superconducting phases, taking into account the non-zero current quark mass.}

\author{Tomasz L. Partyka}

\affiliation{Smoluchowski Institute of Physics, Jagellonian
University, Reymonta 4, 30-059 Krak\'ow, Poland}

\begin{abstract}
We have shown, that the possibility of the existence of the mixed phase of the non-uniform chiral (NCh) and the color superconducting (2SC) ground state depends significantly on the choice of the parameters and type of the regularization scheme. Our calculations indicates, that in the 3d cut-off regularization scheme, the mixed region of the NCh and the 2SC phases exists for a broad set of NJL model parameters. However, in the Schwinger regularization scheme, if parameters are set to the vacuum values of $f_{\pi}$, $m_{\pi}$ and $<\bar{q}q>$, then, the mixed region of the NCh and the 2SC phases does not exists.
\end{abstract}

\maketitle

\section{Introduction}
\mdseries
The goal of this study is to examine the influence of the current quark mass on the phase structure of the strongly interacting quark matter.
The part of the QCD phase diagram that was attracting peoples attention for many years is a region of relatively low temperatures and moderate baryon densities.
In this region, a variety of phases are possible. Especially, a competition between the chiral condensate and the diquark condensate takes place  \cite{shuryak_alford_son} and both these condensates might coexist. Moreover, since a spatially non-uniform chiral ground state was suggested \cite{dautry_bron_sad_jap}, the situation is even more complicated. Many studies was devoted to this issue and most of them was done in the chiral limit. Taking into account both a non-zero current quark mass and a chiral density wave ansatz, makes calculation difficult. The interesting method to deal with that mathematical inconvenience was suggested recently \cite{maedan1, maedan}. The clue is to expand the quark propagator in a perturbative series. We will follow that method, working with the Nambu - Jona-Lasinio model, with two flavors and three colors. It is now well known, that within the mean field approximation in the chiral limit, the chiral density wave ansatz guarantees a lower minimum (in the moderate baryon density region) than a uniform chiral phase. But what happens if we include the non-zero current quark mass? In Ref. \cite{maedan1} it was shown, that in the Schwinger regularization at zero temperature, introduction of the non-zero current quark mass does not exclude the possibility of existence of the spatially inhomogeneous ground state. However, there are open questions. First of all, whether, away from the chiral limit, the non-uniform chiral phase manifests also in different types of regularizations. This is a crucial aspect, especially when doing model dependent calculations \cite{tlms}. Secondly, there is a question, how the non-zero temperature does influence the phase diagram? Finally, we would like to resolve whether there is a possibility of the existence of the mixed region of the 2SC color superconducting and the non-uniform chiral phases. \\

\section{Model} 
\mdseries
The starting point is based on the Nambu - Jona-Lasinio model with two flavours and three colors.
Because we will often refer to Ref. \cite{maedan1}, we must notice that in the present case, the initial lagrangian density differs by adding a charge conjugate fields $\psi_{C}$ and by including a diquark terms.
\begin{eqnarray}
&&H=\int_x[\bar{\psi}(i\gamma^\nu\partial_\nu +\mu\gamma_0-m)\psi+
\bar{\psi}_{C}(i\gamma^\nu\partial_\nu +\mu\gamma_0-m)\psi_{C}\nonumber \\
&&+G\left[ (\bar{\psi}\psi )^2+(\bar{\psi}i\gamma_5\vec{\tau}\psi )^2\right]
+G^\prime (\bar{\psi}_ci\gamma_5\tau_2\lambda^A\psi ) (\bar{\psi}i\gamma_5\tau_2\lambda^A\psi_c )],
\end{eqnarray}
where $\psi $ is the quark field, $\psi_C=C\bar{\psi}^T$ is the conjugate field
and $\mu $ is the quark chemical potential. We assume that current quark masses are equal, $m_{u}=m_{d}=m$. The value of the mass $m$ is fitted with the other parameters of the model. The color, flavor and spinor indices are suppressed. The vector
$\vec{\tau}$ is the isospin vector of Pauli matrices and $\lambda^A$, $A=2,5,7$ are three color
antisymmetric $SU(3)$ group generators. The integration $\int_x=\int_0^\beta d\tau\int d^3x $,
where $\beta $ is the inverse temperature and the derivative operator is $\partial_\nu = (i\partial_\tau ,\vec{\nabla})$.
Two coupling constants $G$ and $G^\prime $ describe interactions which are responsible for the creation of
quark-antiquark and quark-quark condensates respectively. Both couplings are treated as independent.
There is also an additional parameter $\Lambda$ which defines the energy scale below which
the effective theory applies. It is introduced through the regularization procedure.
We are working in the mean field approximation within the ansatz \cite{sad1}
\begin{eqnarray}
\langle\bar{\psi}\psi\rangle = -\frac{M}{2G}\cos\vec{q}\cdot\vec{x},\;\;\;
\langle\bar{\psi}i\gamma_5\tau^a\psi\rangle = -\frac{M}{2G}\delta_{a3}\sin\vec{q}\cdot\vec{x},\;\;\;
\langle\psi\tau_2\lambda^AC\gamma_5\psi\rangle = \frac{\Delta}{2G^\prime}\delta_{A2}
\end{eqnarray}
which describes three possible phases: the chiral uniform phase Ch ($\vec{q}=0, M\neq 0, \Delta =0$),
the non-uniform chiral phase NCh ($\vec{q}\neq 0, M\neq 0, \Delta =0$) and the color superconducting
phase 2SC ($\vec{q} = 0, M = 0, \Delta\neq 0$). Introduction of the non-zero current quark mass results in a fact that a chiral phase order parameter is always larger than the mass $m$, and the chiral symmetry is only partially restored. The presence of the non-zero current mass, causes also, that the 2SC phase always coexists with a  small quark condensate.
We will perform the calculations in a Nambu-Gorkov basis $\chi^{T}=(\psi,\psi_{C})$ \cite{huang}.
To deal with the dependence of the quark propagator on the space coordinates, we make a standard chiral rotation. The mean-field partition function has a form
\begin{eqnarray}
Z_{MF}=\int D\overline{\chi}D\chi\ \exp{H_{MF}},
\end{eqnarray}
\begin{eqnarray}
H_{MF}=\frac{1}{2}\int_{0}^{\beta}\int{d^{3}x}\ {\biggl[\overline{\chi}S^{-1}\chi-\frac{(M_{t}-m)^{2}}{2G}-\frac{\left|\Delta\right|^{2}}{2G'}\biggr]},
\end{eqnarray}
where
\begin{eqnarray}
\label{5}
S^{-1}_{0}=\left[
\begin{array}{c c} i\gamma^{\nu}(\partial_{\nu}-\frac{1}{2}i\gamma_{5}\tau_{3}q^{\nu})+\mu\gamma_{0}-M_{t} & i\gamma_{5}\tau_{2}\lambda_{2}\Delta^{*} \\ i\gamma_{5}\tau_{2}\lambda_{2}\Delta & i\gamma^{\nu}(\partial_{\nu}-\frac{1}{2}i\gamma_{5}\tau_{3}q^{\nu})-\mu\gamma_{0}-M_{t}]
\end{array}
\right],
\end{eqnarray}
 
\begin{eqnarray} 
V_{m}=\left[
\begin{array}{c c} m[\exp(-i\gamma_{5}\tau_{3}\vec{q}\cdot\vec{x})-1] & 0 \\0 & m[\exp(-i\gamma_{5}\tau_{3}\vec{q}\cdot\vec{x})-1] 
\end{array}
\right]
\end{eqnarray}
\space
and
\begin{eqnarray}
q^{\nu} = (0,\vec{q}),\;\; S^{-1}=S_{0}^{-1}-V_{m},\;\; M_{t}=M+m.
\end{eqnarray}
\space
To determine the ground state of the system we calculate the thermodynamic potential
\be
\Omega=\frac{(M_{t}-m)^{2}}{4G}+\frac{\left|\Delta\right|^{2}}{4G'}-\frac{T}{V}\ln\left\{\int D\overline{\chi}D\chi\ \exp\biggl(\frac{1}{2}\int_{0}^{\beta}\int{d^{3}x}\ {\overline{\chi}S^{-1}\chi}\biggr)\right\}.
\ee
Expanding the termodynamic potential up to the terms of order $V_{m}$ we decompose $\Omega$ into the two parts \cite{maedan1} 
\begin{eqnarray}
\Omega=\Omega_{0}+\delta\Omega&=&\frac{(M_{t}-m)^{2}}{4G}+\frac{\left|\Delta\right|^{2}}{4G'}-\frac{T}{V}\ln\left\{\int D\overline{\chi}D\chi\ \exp\biggl(\frac{1}{2}\int_{0}^{\beta}\int{d^{3}x}\ {\overline{\chi}S_{0}^{-1}\chi}\biggr)\right\}\nonumber \\
&+&\frac{T}{V}\;\frac{\int D\overline{\chi}D\chi\ (\int_{0}^{\beta}\int{d^{3}x}\ {\overline{\chi}\frac{V_{m}}{2}\chi}) \exp(\frac{1}{2}\int_{0}^{\beta}\int{d^{3}x}\ {\overline{\chi}S_{0}^{-1}\chi})}{\int D\overline{\chi}D\chi\ \exp(\frac{1}{2}\int_{0}^{\beta}\int{d^{3}x}\ {\overline{\chi}S_{0}^{-1}\chi})}+O(V_{m}^{2}).
\end{eqnarray}
Using the standard method one can calculate the thermodynamic potential $\Omega_{0}$ \cite{sad}
\begin{eqnarray}
&&\Omega_{0}= \frac{(M_{t}-m)^2}{4G} +\frac{|\Delta |^2}{4G^\prime}+2\sum_{s=\pm}\int_{E_s\leq\mu}\frac{d^3k}{(2\pi)^3} (E_s-\mu)
-2\sum_{s=\pm}\int\frac{d^3k}{(2\pi)^3} \biggl(E_s+\sum_{i=\pm}E^\Delta_{i,s}\biggr)-\nonumber \\
&&\sum_{s=\pm}4T\int\frac{d^{3}k}{(2\pi)^{3}}\left[\ln\left(1+\exp \biggl(-\frac{\sqrt{(E{s}+\mu )^{2}+|\Delta|^{2}}}{T}\biggr)\right)+\ln\left(1+\exp\biggl(-\frac{\sqrt{(E{s}-\mu)^{2}+|\Delta|^{2}}}{T}\biggl)\right)\right]-\nonumber \\
&&\sum_{s=\pm}2T\int\frac{d^{3}k}{(2\pi)^{3}}\left[\ln\left(1+\exp\biggl(-\frac{(E{s}+\mu)}{T}\biggr)\right)+\ln\left(1+\exp\biggl(-\frac{|E{s}-\mu|}{T}\biggr)\right)\right]
\end{eqnarray}
where
\be
\label{11}
E^\Delta_{\pm,s}=\sqrt{(\mu\pm E_s)^2+|\Delta|^2},\;\;
E_\pm=\sqrt{\vec{k}^2+M_{t}^2+\frac{\vec{q}^{\,2}}{4}\pm \sqrt{(\vec{q}\cdot\vec{k})^2+M_{t}^2\vec{q}^{\,2}}}.
\vspace{0.5 cm}
\ee
The term $\delta\Omega$ gives a non-zero contribution to the potential only if $q\neq0$, and in a momentum-frequency basis, has a form \cite{maedan1}

\be
\label{12}
\delta\Omega=-m T\sum_{n=-\infty}^{n=+\infty}\int\frac{d^{3}k}{(2\pi)^{3}}\;\frac{1}{2}\; tr S_{0}.
\ee
The result of Matsubara sum is rather complicated and more detailed study can be found in the Appendix
\begin{eqnarray}
\label{13}
\delta\Omega&=&-\;2m\sum_{s=\pm}\int\frac{d^{3}k}{(2\pi)^{3}}\biggl[-\frac{1}{2}\frac{\partial}{\partial M_{t}}(E_{s}+E^{\Delta}_{+,s})\biggl(1+\exp(-\beta E^{\Delta}_{+,s})\biggr)^{-1}\nonumber \\
&&-\;\frac{1}{2}\frac{\partial}{\partial M_{t}}(E_{s}-E^{\Delta}_{+,s})\biggl(1+\exp(\beta E^{\Delta}_{+,s})\biggr)^{-1}+\;\frac{1}{2}\frac{\partial}{\partial M_{t}}(E_{s}-E^{\Delta}_{-,s})\biggl(1+\exp(-\beta E^{\Delta}_{-,s})\biggr)^{-1}\nonumber \\&&+\;\frac{1}{2}\frac{\partial}{\partial M_{t}}(E_{s}+E^{\Delta}_{-,s})\biggl(1+\exp(\beta E^{\Delta}_{-,s})\biggr)^{-1}\biggr]\nonumber\\
&&-\;m\sum_{s=\pm}\int\frac{d^{3}k}{(2\pi)^{3}}\biggl[-\frac{\partial}{\partial M_{t}}(E_{s})\biggl(1+\exp(-\beta (E_{s}+\mu))\biggr)^{-1}\nonumber \\
&&+\;\frac{\partial}{\partial M_{t}}(E_{s})\biggl(1+\exp(\beta (E_{s}-\mu))\biggr)^{-1}\biggr].
\end{eqnarray}
\section{Regularization procedure}
\mdseries
The Nambu - Jona-Lasinio model is a nonrenormalizable theory. That is why, the regularization procedure is needed to clearly define the model. We use the two types of regularization procedure: 3d cutoff and the Schwinger (proper time) scheme.
The temperature dependent part of the $\Omega_{0}$ is finite, but the zero temperature limit of the potential $\Omega_{0}$ contains divergent terms. We rewrite $\Omega_{0}(T=0)$ in a form that is more convenient to regularize
\begin{eqnarray}
\label{14}
&&\Omega_{0}(T=0)=\frac{(M_{t}-m)^2}{4G} +\frac{|\Delta |^2}{4G^\prime}
-2\sum_{i=\pm}\int\frac{d^3k}{(2\pi)^3} \left(\sum_{s=\pm}(E^\Delta_{i,s}-E_{s})-2(E^\Delta_{i,0}-E^{\Delta=0}_{i,0})\right)\nonumber \\
&&+\; 2\sum_{s=\pm}\int_{E_s\leq\mu}\frac{d^3k}{(2\pi)^3} (E_s-\mu)
-4\sum_{i=\pm}\int\frac{d^3k}{(2\pi)^3} \left(E^\Delta_{i,0}-E^{\Delta=0}_{i,0}\right)
-6\sum_{s=\pm}\int\frac{d^3k}{(2\pi)^3}(E_s)
\end{eqnarray}
where
\begin{eqnarray}
E^\Delta_{\pm,0}&=&\sqrt{(\mu\pm E_0)^2+|\Delta|^2},\;\;E_0=\sqrt{\vec{k}^2+M_{t}^2} .
\end{eqnarray}
Only the last two terms of equation (\ref{14}) are divergent.
The infinite contribution to the $\delta \Omega$ comes from the following terms
\begin{eqnarray}
&&\sum_{s=\pm}\int\frac{d^{3}k}{(2\pi)^{3}}\biggl[-\frac{1}{2}\frac{\partial}{\partial M_{t}}(E_{s}+E^{\Delta}_{+,s})\left(1+\exp(-\beta E^{\Delta}_{+,s})\right)^{-1}\nonumber \\
&&+\;\frac{1}{2}\frac{\partial}{\partial M_{t}}(E_{s}-E^{\Delta}_{-,s})\left(1+\exp(-\beta E^{\Delta}_{-,s})\right)^{-1}\biggr]
\end{eqnarray}
and
\begin{eqnarray}
\label{17}
\sum_{s=\pm}\int\frac{d^{3}k}{(2\pi)^{3}}\biggl[-\frac{\partial}{\partial M_{t}}(E_{s})\biggl(1+\exp(-\beta (E_{s}+\mu))\biggr)^{-1}\biggr].
\end{eqnarray} 
\begin{itemize}
\item
\vspace{1cm}
\bf{3d cutoff regularization}\\
\mdseries
We regularize divergent integrals by introducing a cutoff $\Lambda$, $\vec{k}^{2}<\Lambda$.\\
\item
\bf{Schwinger (proper time) regularization}\\
\mdseries
This method is more subtle \cite{nakano}. At first, we decompose divergent integrals into the terms that depends on temperature (these terms are finite) and the remaining part. The finite terms contains factors of a type $(1+\exp(-\beta E^{\Delta}_{\pm,s}))^{-1}$. With momentum tending to the infinity, above mentioned factors tends to 1. In the example, with this method, integral (\ref{17}) has a form
\begin{eqnarray}
\label{18}
&&\sum_{s=\pm}\int\frac{d^{3}k}{(2\pi)^{3}}\biggl[-\frac{\partial}{\partial M_{t}}(E_{s})\left(\biggl(1+\exp(-\beta (E_{s}+\mu))\biggr)^{-1}-1\right)\biggr] \nonumber\\
&&+ \int\frac{d^{3}k}{(2\pi)^{3}}\biggl[-\frac{\partial}{\partial M_{t}}(E_{s})\biggr].
\end{eqnarray} 
The last term of equation (\ref{18}) is finally regularized in the Schwinger scheme
\begin{eqnarray}
&&\int\frac{d^{3}k}{(2\pi)^{3}}\biggl[-\frac{\partial}{\partial M_{t}}(E_{s})\biggr]=-\int\frac{d^{4}k}{(2\pi)^{4}}\int_{1/\Lambda^{2}}^{\infty}\frac{d\tau}{\tau}\biggl[-\frac{\partial}{\partial M_{t}}\exp(-\tau(k_{0}^{2}+E_{s}^{2}))\biggr].
\end{eqnarray}
\end{itemize}

\hrulefill

\begin{table}[h]
\begin{center}
\begin{tabular}{ccccc}
          & 3d & S  \\
\hline\hline 
$\Lambda\;\;\;$    & 0.653  & 1.086 \\
$G\Lambda^2\;\;\;$ & 2.1   & 3.68   \\
$M_0\;\;\;$        & 0.313  & 0.2   \\
$m\;\;\;$        & 0.005  & 0.0049   \\
$(-<\bar{u}u>)^{\frac{1}{3}}$       &0.25 &0.25  \\
\hline\hline
\end{tabular}
\end{center} 
\caption{Numerical values of the regularization parameters in GeV (type 1)}
\end{table}
\begin{table}[h]
\begin{center}
\begin{tabular}{ccccc}
          & 3d & S  \\
\hline\hline 
$\Lambda\;\;\;$    & 0.619  & 0.635 \\
$G\Lambda^2\;\;\;$ & 2.24   & 7.53   \\
$M_0\;\;\;$        & 0.35  & 0.38   \\
$m\;\;\;$        & 0.0053  & 0.0153   \\
$(-<\bar{u}u>)^{\frac{1}{3}}$       &0.245 &0.17  \\
\hline\hline 
\end{tabular}
\end{center} 
\caption{Numerical values of the regularization parameters in GeV (type 2)}
\end{table}

\section{Results} 
\mdseries
The parameters of the NJL model are $G, G', \Lambda$ and $m$. As we shall see, the particular choice of the above mentioned parameters can have a decisive influence on the conclusions following from the NJL model. The physical quantities in the vacuum that are used to fix $G, \Lambda$ and $m$ are: $f_{\pi}=$ 93 MeV, $m_{\pi}=$ 135 MeV and $<\bar{u}u>=<\bar{d}d>=$ -(250 $\pm$ 50 MeV)$^{3}$ \cite{klev}. Because of the wide experimental range for the value of the quark condensate density, one can neglect $<\bar{q}q>$ and then, $\Lambda$ becomes a free parameter \cite{loops}. To make the discussion clear, we define the parameters of type 1 (when the $<\bar{q}q>$ value is taken into consideration) and of type 2 (when $\Lambda$ is a free parameter). The fitted parameters are collected in the Tables I, II, where also  the value of the constituent quark mass in the vacuum $M_{0}$ is given.
Coupling constant $G'$ can not be related to any known physical quantity. The larger the value of $G'$ the lower the value of the $\Omega_{0}$ in the minimum corresponding to the 2SC superconductor. Because we want to resolve whether the mixed region of the NCh/2SC phase occurs when $m\neq$ 0, we set $G'=$ 0.5 $G$. That is a lower bound for the coupling constant $G'$ considered among the vast literature on the color superconductivity \cite{berjag, tlms}.
With the fixed parameters $G, G', \Lambda$ and $m$, the thermodynamical potential $\Omega$ is a function of $M_{t}, q, |\Delta|$, chemical potential $\mu$ and temperature T. We begin ours analysis with a zero temperature limit. For a constant $\mu$ and T, we numerically determine a global minimum of $\Omega$ with respect to the free parameters. We restrict our analysis to the region of $\mu$, where numerically obtained $M_{t}$ is greater than mass $m$.

\subsection{3d cutoff regularization}
\begin{figure}[h]
	\centering
		\includegraphics[width=0.55\textwidth]{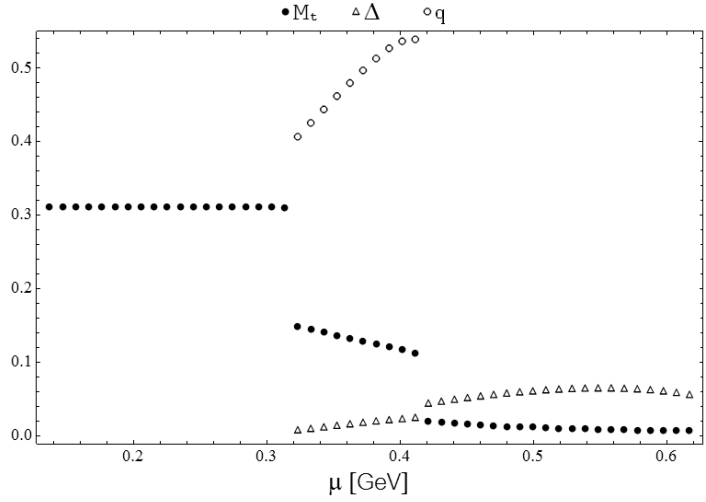}
		\caption{The values of $M_{t}$, q and $\Delta$ in the 3d cut-off regularization scheme for
 T=0 (GeV units). The global minimum related to the mixed phase NCh/2SC is plotted. (type 1 parameters)} \label{number}
\end{figure}
\begin{figure}[h]
	\centering
		\includegraphics[width=0.55\textwidth]{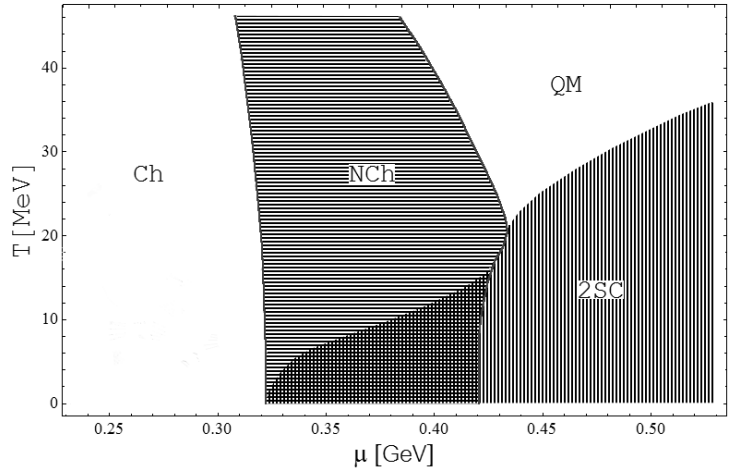}
		\caption{The phase diagram in the $\mu$ - T plane. (type 1 parameters)} \label{number}
\end{figure}
\subsubsection{Type 1 parameters}
In the Fig. 1, dependences of $M_{t}, q$ and $|\Delta|$ on $\mu$ at T = 0 are shown. The spatially inhomogeneous quark phase does exist. There is a global minimum corresponding to the mixed phase of non-uniform chiral (NCh) and color superconducting (2SC) phases. At zero temperature, its value is slightly lower than the value of the local minimum corresponding to the mixed phse of 2SC and uniform chiral (Ch) phases, in the interval of $\mu$ from 0.322 GeV to 0.420 GeV (Fig. 1). As we can expected \cite{sad}, non-zero temperature does not favours a non-uniform condensate. The phase diagram in the $\mu$ - T plane is shown in Fig. 2. For temperatures between 0 and 20 MeV, the interval of existence of the inhomogeneous phase is practically constant, but the region of a coexistence of NCh and 2SC phases is getting smaller with growing temperature (Fig. 2, 3). Above T = 16 MeV, spatially inhomogeneous quark condensate exists  without a diquark condensate, and region in which the global minimum is associated to NCh is systematically decreasing.
\begin{figure}[h]
	\centering
		\includegraphics[width=0.55\textwidth]{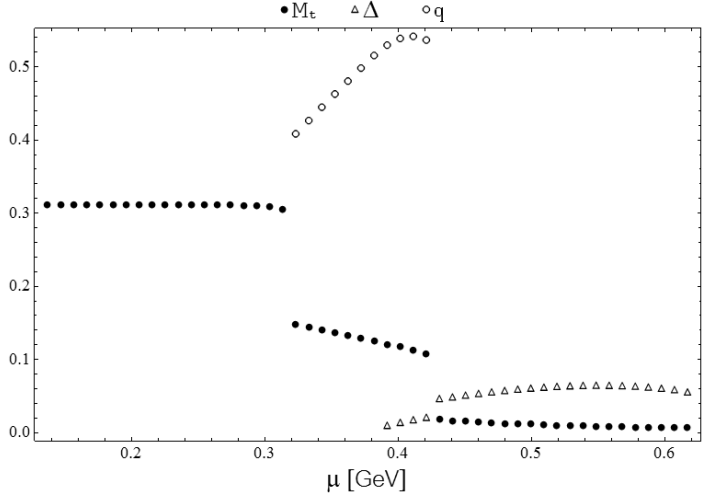}
		\caption{The values of $M_{t}$, q and $\Delta$ in the 3d cut-off regularization scheme for T=0.011 (GeV units, type 1 parameters).} \label{number}
\end{figure}
\subsubsection{Type 2 parameters}
\begin{figure}[h]
	\centering
		\includegraphics[width=0.55\textwidth]{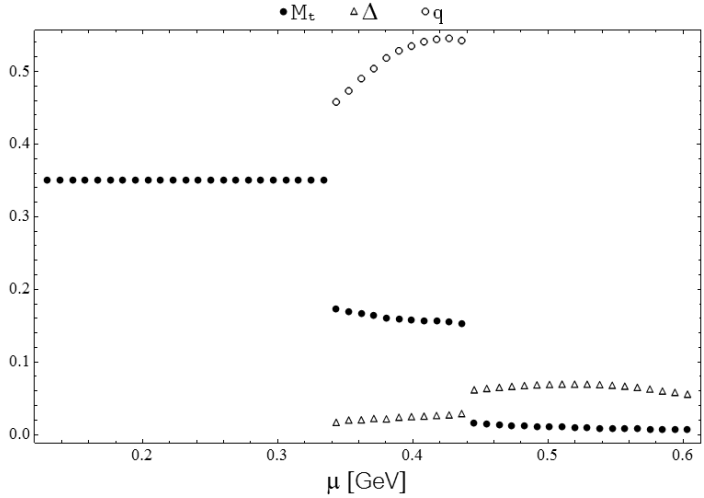}
		\caption{The values of $M_{t}$, q and $\Delta$ in the 3d cut-off regularization scheme for T=0 (GeV units). The global minimum related to the mixed phase NCh/2SC is plotted. (type 2 parameters)} \label{number}
\end{figure}
The wide experimental range for the quark condensate density allows us to shift the value
of $\Lambda$. Moving $\Lambda$ above the previous value 653 MeV results in a decrease of a value of the coupling
constant G. For the lower values of G, minimum associated to the NCh phase
is getting more shallow, that case will also be discussed later. Hence, we will rather reduce the value of $\Lambda$.  On the other hand, $\Lambda$ is an upper boundary for the energy scale, so the possible interval for the values of $\Lambda$ is narrow. We set $G\Lambda^{2}=$ 2.44, $\Lambda=$ 619 MeV, $M_{0}=$ 350 MeV and $m=$ 5.3 MeV. With this choice of parameters, $(-\frac{1}{2}<\bar{q}q>)^{\frac{1}{3}}=245$ MeV and it is still within the experimental range for the value of the quark condensate density. Fig. 4 presents the dependences of $M_{t}, q$ and $|\Delta|$ on $\mu$ at T = 0. The spatially inhomogeneous quark phase exists in a region of $\mu$ from 340 to 440 MeV. In that region, the 2SC color superconducting phase coexists with the NCh phase. As one can expected, with growing T, above mentioned region is getting narrower and the value of the color superconducting order parameter $|\Delta|$ is decreasing (Fig. 5). Above T = 20 MeV, there is no mixed region of the NCh/2SC phase and these phases exists separately.
\begin{figure}[h]
	\centering
		\includegraphics[width=0.55\textwidth]{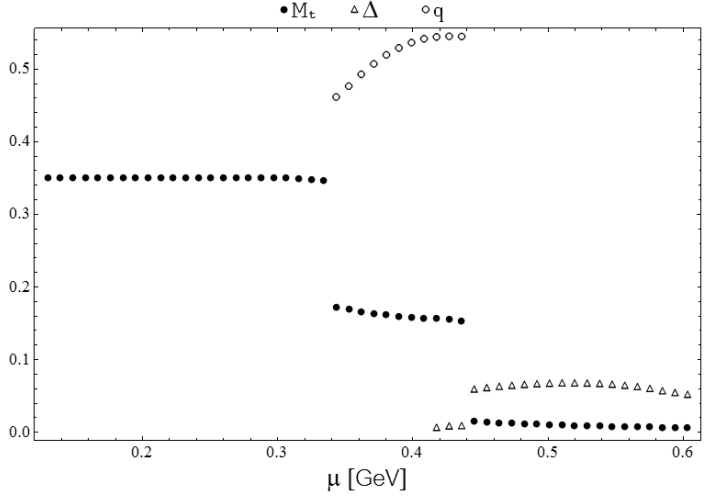}
		\caption{The values of $M_{t}$, q and $\Delta$ in the 3d cut-off regularization scheme for T=0.016 (GeV units). The global minimum related to the non-uniform chiral solution is plotted.} \label{number}
\end{figure}
\subsection{Schwinger regularization}
\begin{figure}[h]
	\centering 
		\includegraphics[width=0.55\textwidth]{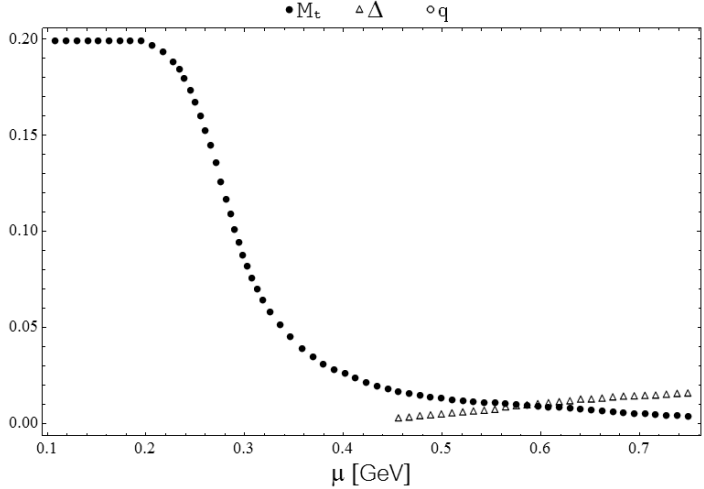}
		\caption{The values of $M_{t}$ and $\Delta$ in the Schwinger regularization scheme for T=0 (GeV units).} \label{number}
\end{figure}
\subsubsection{Type 1 parameters}
In the proper time scheme at T = 0, the mixed phase of the NCh and 2SC phases is only a local minimum, and the values of the vector $q$ corresponding to that local minimum are unphysical ($q$ is much greater than the $\Lambda$ scale). In Fig. 6, there are presented dependences of $M_{t}$ and $|\Delta|$ on $\mu$ at T = 0. Even at T = 0, we observe a smooth dependence of $M_{t}$ on a chemical potential. $M_{t}$ decreases from its vacuum value, across the narrow range of the rapid variability, down to the 5 MeV. In this regularization scheme, the values of $|\Delta|$ are of order of the magnitude smaller than in the cutoff scheme. That fact was already observed \cite{tlms}. For the values of T above 5 MeV, the diquark condensate vanishes.\\
 \subsubsection{Type 2 parameters}
The limit values of the NJL parameters for which $(-\frac{1}{2}<\bar{q}q>)^{\frac{1}{3}}>200$ MeV are $G\Lambda^{2}=$ 4.58, $\Lambda=$ 760 MeV and $m=$ 12.6 MeV. However, at these limit values, there is no non-uniform phase. Guided by the findings of \cite{maedan1}, we set $G\Lambda^{2}=$ 7.53, $\Lambda=$ 635 MeV, $M_{0}=$ 380 MeV and $m=$ 15.3 MeV. With this choice of parameters, $(-\frac{1}{2}<\bar{q}q>)^{\frac{1}{3}}=170$ MeV and it is already beyond the experimental range for the value of the quark condensate. As can be seen in Fig. 7, the mixed region of the non-uniform chiral/2SC superconductor ground state appears from $\mu=$ 355 MeV. Above the value of quark chemical potential equal 450 MeV, the value of the wave vector q is greater than $\Lambda$. Therefore, above these limit, discussed model is unphysical.
\begin{figure}[h]
	\centering
		\includegraphics[width=0.55\textwidth]{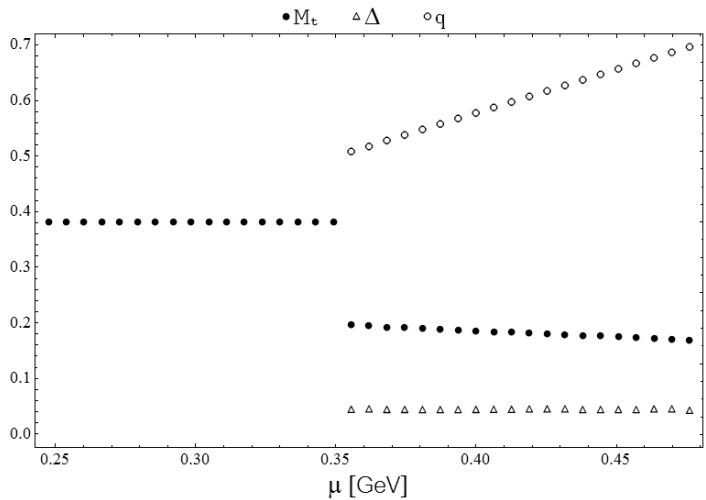}
		\caption{The values of $M_{t}$, q and $\Delta$ in the Schwinger regularization scheme for T=0 (GeV units). $G\Lambda^{2}$=7.53, $\Lambda$=0.635 GeV, m=0.0153 GeV} \label{number}
\end{figure}
\subsection{Discussion}
To compare the behavior of the potential $\Omega$ in  different regularizations and at the different choice of parameters, we take the factor $\Lambda^{4}$ ahead of the $\Omega$ and then we work with the dimensionless quantities.
The $\delta \Omega$ contribution to the thermodynamic potential is proportional to the ratio $m/\Lambda$. When the value of q is non-zero, $\delta \Omega$ is positive and therefore, the value of $\Omega$ is higher than the value of $\Omega_{0}$. As a consequence, $\delta \Omega$ term does not favour the possibility of the existence of the NCh phase. One might ask a question, why so small mass $m$ (in comparison with other energy scales) has so significant influence on the $\Omega$. In the 3d cutoff scheme with the parameters of type 1, the ratio $m/\Lambda$ = 0.0076  and with type 2, $m/\Lambda$ = 0.0085. Similarly, in the proper time scheme, with the parameters of type 1, the ratio $m/\Lambda$ = 0.0045 and with type 2, $m/\Lambda$ = 0.024. It appears, that the difference between the value of the $\Omega_{0}$ in the global minimum corresponding to the NCh/2SC phase and the value of the $\Omega_{0}$ in the local minimum corresponding to a uniform chiral phase is also of order of a few percent. 
Therefore, even so small mass $m$ can moves the global minimum.
There is also another question, why the inhomogeneous ground state solution
appears for the rather higher current masses. Of course, the parameters of the NJL model are linked with each other. It turns, that a larger $m$ requires a larger coupling constant G. 
With increasing G, the distance between the value of the $\Omega_{0}$ in the global minimum corresponding to the NCh/2SC phase and the value of the $\Omega_{0}$ in the local minimum corresponding to a uniform chiral phase is more distinctive. In Fig. 8. we show the behavior of the values of the full potential $\Omega$ in the minima corresponding to the NCh/2SC and Ch/2SC solutions as a function of mass $m$ in the 3d cutoff scheme. The upper limit on the value of mass $m$ is $m\approx$ 5.45 MeV when the vacuum value of $M_{0}$ exceeds 400 MeV. On the other hand, below the value of mass $m\approx$ 4.7 MeV the non-uniform solution disappears.
\begin{figure}[t]
	\centering
		\includegraphics[width=0.55\textwidth]{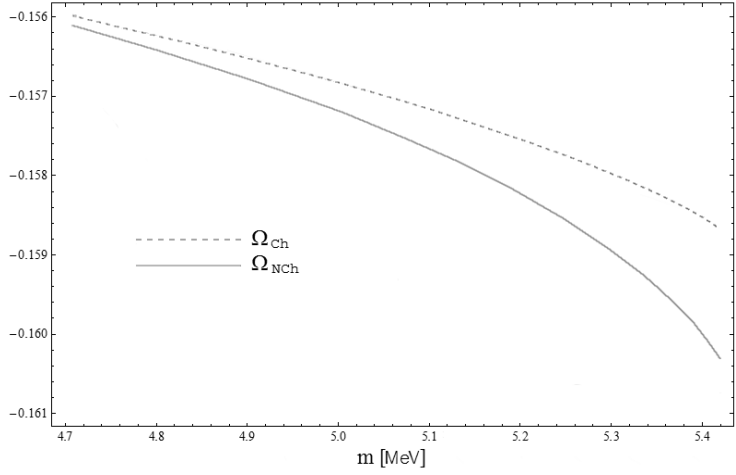}
		\caption{The values of the $\Omega$ in the minima corresponding to the NCh/2SC phase
and to the Ch/2SC phase (dashed line), as a function of mass $m$ in $\mu=$ 360 MeV (3d cut-off, type 1 parameters)} \label{number}
\end{figure}
\section{Conclusions}
We have shown, that the possibility of the existence of the mixed phase of the non-uniform chiral (NCh) and color superconducting (2SC) ground state depends significantly on the choice of the parameters and type of the regularization scheme. Our calculations indicates, that in the 3d cut-off regularization scheme, the mixed region of the NCh and 2SC phases exists for a broad set of NJL model parameters. However, in the Schwinger regularization scheme, if parameters are set to the vacuum values of $f_{\pi}$, $m_{\pi}$ and $<\bar{q}q>$, then, the mixed region of the NCh and 2SC phases does not exists. In the moderate baryon density region, there is a local minimum associated to the NCh and 2SC phases, but the ground state is realized by a chiral uniform, or in a small range of temperatures above zero by the uniform chiral (Ch) and superconductor (2SC) phases. If we release the constrain on the parameters coming from the value of $<\bar{q}q>$, then, the NCh solution might be an absolute minimum of the thermodynamical potential in the moderate baryon density region. 
The result, that is irrespective for the parameters and regularization method choice is that, if for some chemical potential (at T = 0) there is a minimum associated to the NCh/2SC phase, its value is always lower than the value of the minimum associated to the spatially inhomogeneous quark condensate with $|\Delta|$ equal to zero. Because we chose $G'$= 0.5 $G$, that is a lower bound for the estimated value of a coupling $G'$, it can be assumed, that if the NCh condensate exists at zero temperature, it exists together with a diquark condensate. With growing temperature, a diquark condensate starts up at a greater $\mu$ and the region of a coexistence of the NCh and 2SC phases disappears.
\section{Appendix}
\mdseries
In the expression (\ref{12}) 
\begin{eqnarray}
 \int\frac{d^{3}k}{(2\pi)^{3}}T\sum_{n=-\infty}^{n=+\infty}\;\frac{1}{2}\; tr S_{0}\nonumber
\end{eqnarray}
the trace is taken over the Nambu-Gorkov, color and flavor indices. The sum is taken over the Matsubara frequencies.
The diagonal elements of the $S_{0}$ operator Eq. (\ref{5}) in the momentum frequency bassis have a form

\begin{eqnarray} S_{0_{\alpha\beta\xi}}&=&\gamma_{0}\biggl(|\Delta|^{2}\left(-(p_{0}+(-1)^{\xi}\mu)+\gamma_{0}(\vec{k}\vec{\gamma}-M_{t}+(-1)^{\alpha}(\delta^{\beta}_{2}\delta^{3-\beta}_{1}+\delta^{\beta}_{1}\delta^{3-\beta}_{2})\;\frac{q}{2}\gamma_{5}\gamma_{3})\right)^{-1} \nonumber \\ &+&(p_{0}-(-1)^{\xi}\mu)+\gamma_{0}(\vec{k}\vec{\gamma}-M_{t}+(-1)^{\alpha}(\delta^{\beta}_{2}\delta^{3-\beta}_{1}+\delta^{\beta}_{1}\delta^{3-\beta}_{2})\;\frac{q}{2}\gamma_{5}\gamma_{3})\biggr)^{-1},
\end{eqnarray}
where indices $\alpha, \xi$ refers respectively to the flavor and Nambu-Gorkov space and are equal 1, 2. Index $\beta$ refers to the color space and takes the values 1, 2, 3.
We calculate that trace in the eigenbasis of a matrix $\gamma_{0}(\vec{k}\vec{\gamma}-M_{t}\pm\frac{q}{2}\gamma_{5}\gamma_{3})$. The eigenvalues of a matrix $\gamma_{0}(\vec{k}\vec{\gamma}-M_{t}\pm\frac{q}{2}\gamma_{5}\gamma_{3})$ to the eigenvectors $\psi_{i}$ are $\lambda_{i}=\pm E_{\pm}$ (\ref{11}). For example, the trace of a one of these diagonal elemsnts is
\begin{eqnarray}
tr S_{0_{111}}=\sum_{i}\psi^{+}_{i}\gamma_{0}\psi_{i}\left(|\Delta|^{2}(-p_{0}+\mu+\lambda_{i})^{-1}+(p_{0}+\mu+\lambda_{i})\right)^{-1}.
\end{eqnarray}
Finally we get
\begin{eqnarray}
\label{22}
&&4M_{t}\;\frac{E_{+}^{2}-\vec{p}^{\; 2}-M_{t}^{2}+\frac{q^{2}}{4}}{2(E_{+}^{2}-E_{-}^{2})}\;\biggl[-\frac{1}{E_{+}}\frac{p_{0}-\mu+E_{+}}{(p_{0}-E^{\Delta}_{-+})(p_{0}+E^{\Delta}_{-+})} +\frac{1}{E_{+}}\frac{p_{0}-\mu-E_{+}}{(p_{0}-E^{\Delta}_{++})(p_{0}+E^{\Delta}_{++})}\;\biggr]
\nonumber \\
&+&4M_{t}\;\frac{E_{-}^{2}-\vec{p}^{\; 2}-M_{t}^{2}+\frac{q^{2}}{4}}{2(E_{+}^{2}-E_{-}^{2})}\;\biggl[ -\frac{1}{E_{-}}\frac{p_{0}-\mu+E_{-}}{(p_{0}-E^{\Delta}_{--})(p_{0}+E^{\Delta}_{--})}
+\frac{1}{E_{-}}\frac{p_{0}-\mu-E_{-}}{(p_{0}-E^{\Delta}_{+-})(p_{0}+E^{\Delta}_{+-})}\;
\biggr].\nonumber \\
\end{eqnarray}
Expression (\ref{13}), is obtained by the contour integration of expression (\ref{22}) \cite{maedan1} and by a summation over the whole set of indices.
\vspace{0.5 cm}
 
\textbf{Acknowledgement:}
I would like to thank Professor Mariusz Sadzikowski for many interesting discussions and comments.

\end{document}